\documentclass[%
 aip,
 amsmath,amssymb,
 reprint,%
]{revtex4-1}

\usepackage{graphicx}
\usepackage{dcolumn}
\usepackage{bm}

\usepackage[utf8]{inputenc}
\usepackage[T1]{fontenc}
\usepackage{mathptmx}
\usepackage{etoolbox}

\makeatletter
\def\@email#1#2{%
 \endgroup
 \patchcmd{\titleblock@produce}
  {\frontmatter@RRAPformat}
  {\frontmatter@RRAPformat{\produce@RRAP{*#1\href{mailto:#2}{#2}}}\frontmatter@RRAPformat}
  {}{}
}%
\makeatother
\begin{document}

\preprint{AIP/123-QED}

\title{Direct Epitaxial Growth and Deterministic Device Integration of high-quality Telecom O-Band InGaAs Quantum Dots on Silicon Substrate}
	\author{I. Limame}
	\altaffiliation{imad.limame@tu-berlin.de}
	\email{stephan.reitzenstein@physik.tu-berlin.de}
    \affiliation{Institute of Physics and Astronomy, Technical University of Berlin, Hardenbergstraße 36, D-10623 Berlin, Germany}
    \author{P. Ludewig}
    \affiliation{NAsP III/V GmbH, Hans-Meerwein-Straße 6, D-35032 Marburg, Germany}
    \author{A. Koulas-Simos}
	\author{C. C. Palekar}
    \author{J. Donges} 
    \author{C.-W. Shih} 
	\author{K. Gaur}
	\author{S. Tripathi}
        \author{S. Rodt}
   \affiliation{Institute of Physics and Astronomy, Technical University of Berlin, Hardenbergstraße 36, D-10623 Berlin, Germany}
    \author{W. Stolz}
	\affiliation{NAsP III/V GmbH, Hans-Meerwein-Straße 6, D-35032 Marburg, Germany}
    \author{K. Volz}
	\affiliation{mar.quest | Marburg Center for Quantum Materials and Sustainable Technologies, Philipps University Marburg, 35032 Marburg, Germany}
    \affiliation{Department of Physics, Philipps University Marburg, Hans Meerwein Str. 6, 35032 Marburg, Germany}
	\author{S. Reitzenstein}
	\altaffiliation{stephan.reitzenstein@physik.tu-berlin.de}
	\affiliation{Institute of Physics and Astronomy, Technical University of Berlin, Hardenbergstraße 36, D-10623 Berlin, Germany}

\date{\today}

\begin{abstract}
	
Semiconductor quantum dots (QDs) are key building blocks for photonic quantum technologies, enabling practical sources of non-classical light. A central challenge for scalable integration is the direct epitaxial growth of high-quality emitters on industry-compatible silicon platforms. Furthermore, for long-distance fiber-based quantum communication, emission in the telecom O- or C-band is essential. Here, we demonstrate the direct growth of high-quality InGaAs/GaAs QDs emitting in the telecom O-band using a strain-reducing layer approach on silicon. Deterministic integration of individual QDs into circular Bragg grating resonators is achieved via in-situ electron-beam lithography. The resulting devices exhibit strong out-coupling enhancement, with photon extraction efficiencies up to $(40 \pm 2)\%$, in excellent agreement with numerical simulations. These results highlight the high material quality of both the epitaxial platform and the photonic nanostructure, as well as the precise lateral positioning of the emitter within 20~nm of the resonator center. At cryogenic temperature (4~K) and low excitation power ($0.027\times P_\text{sat}$), the devices show excellent single-photon purity, exceeding 99\%. Operation at elevated temperatures of 40~K and 77~K, compatible with compact Stirling cryo-coolers and liquid-nitrogen cooling, reveals robust performance, with single-photon purity maintained at $(88.4 \pm 0.6)\%$ at 77~K. These results demonstrate a practical and scalable route toward silicon-based quantum light sources and provide a promising path for cost-effective fabrication and seamless integration of quantum photonics with classical electronics, representing an important step toward large-scale, chip-based quantum information systems.

\end{abstract}

\maketitle

\section{INTRODUCTION}

Designating 2025 as the International Year of Quantum Science and Technology highlights the evolution of a century of foundational discoveries in quantum physics has evolved into a global race to harness quantum technologies, with non-classical light sources at the forefront in photonic quantum technology developments \cite{Lele2021, Krause2024, Singh2024}. In quantum photonics, semiconductor quantum dots (QDs) are one of the most promising and advanced platforms for generating non-classical light \cite{Shields2007-wl, Senellart2017-mc, Arakawa2020}. Their ability to emit single photons in principle on demand, with high single-photon purity \cite{Schweickert2018}, high indistinguishability \cite{Reitzenstein2025}, and entanglement fidelity \cite{Schimpf2021} spanning the wavelength range of 780 to 1550 nm \cite{Arakawa2020} makes them particularly attractive for various applications, including quantum key distribution \cite{Rau2014-pc, Zhang2025-vo, Yu2023}, photonic quantum computing \cite{Couteau2023-sl, Maring2024-ia}, and quantum sensing \cite{Crawford2021}. Their compatibility with well-established growth techniques, such as molecular beam epitaxy and metal-organic chemical vapor deposition (MOCVD) further enhances their appeal to the scientific community, semiconductor industry and photonics industry.

\begin{figure*} 
	\centering
	\includegraphics[width=0.8\textwidth]{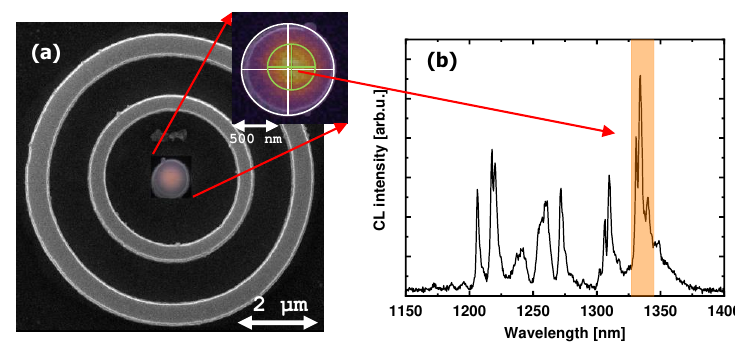}  
	\caption{(a) Scanning electron microscopy (SEM) image of the investigated QD-CBG structure fabricated on a GaP/Si template. A CL intensity map spanning 1326–1345 nm is overlaid at the center of the central mesa. The inset shows a magnified view, where the white circle with cross indicates the geometric center of the mesa, and the green circle with cross marks the fitted position of the integrated quantum emitter. The emitter is offset from the mesa center by only (20 $\pm$ 2) nm. (b) CL spectrum of the QD-CBG indicating the spectral range of the relevant QD. Due to a rather high QD density of $\sim (1-5 \times 10^9$)~cm$^{-2}$ multiple QDs with distinct emission energies were incorporated into the central mesa region of the structure.}
	\label{Fig:CL}
\end{figure*}

Despite the remarkable progress in the development of semiconductor QDs and quantum light sources more broadly, there are still key challenges in terms of scalability, cost-effectiveness, and integration compatibility with the dominant silicon-based electronic and photonic platforms \cite{Bogdanov2016}. Significant advances have been made in post-growth integration techniques, such as wafer bonding and transfer printing—for combining III–V semiconductors with silicon \cite{Katsumi2019, Vijayan2024}. Direct epitaxial growth of III–V QDs on silicon is widely regarded as essential for future quantum photonic integration, offering scalable and monolithic solutions. This approach promises to combine the superior optical properties of III–V materials with the excellent electronic performance and low manufacturing cost of silicon technology. A promising method of achieving direct III–V growth on silicon has been developed, involving the use of a GaP nucleation layer as an intermediate buffer on the silicon substrate \cite{Volz2011}. This buffer is followed by the sequential deposition of GaAs, AlGaAs, and GaInP layers, which reduce strain and defects, effectively suppressing dislocations and lattice mismatch effects \cite{Nmeth2008, Volz2011, Beyer2011, Beyer2013}. Growth templates of this kind enable III–V materials to be grown on silicon with sufficient crystal quality for laser applications\cite{Huang2014, Jung2017, Shang2021}. Building on this platform, we have recently demonstrated the direct growth of high-quality single InGaAs/GaAs QDs on the GaP-based template\cite{Limame2024b}. The reported QDs, which emit in the 930~nm spectral range, exhibit optical and quantum-optical properties, including linewidth and single-photon purity, comparable to reference QDs grown on conventional GaAs substrates.

In this study, we demonstrate the direct growth of InGaAs/GaAs QDs emitting in the telecom O-band on oriented (\( \pm 0.5^\circ \)) Si (001) substrates. In order to enable emission within the important 1.3~\textmu m spectral range for mid-range fiber-based quantum communication, the epitaxial structure incorporates an intermediate GaP buffer layer, followed by multiple layers designed to reduce strain and defect density. Following growth, the QDs are deterministically integrated into circular Bragg grating (CBG) structures using in-situ electron beam lithography (iEBL) \cite{Rodt2021} to enhance the photon extraction efficiency (PEE). The fabricated quantum emitters exhibit excellent optical properties that are comparable to those of state-of-the-art telecom O-band InGaAs/GaAs QDs grown on GaAs substrates\cite{Srocka2020-ms, Engel2023, Alqedra2025-fh}. Under pulsed, non-resonant wetting-layer excitation at 1240~nm and 4~K, we observe a single-photon emission purity of $(99.3 \pm 0.1)\%$ at low excitation strength and $(82.8 \pm 0.4)\%$ at saturation pump power ($P_{sat}$). Single-photon emission with a purity of 88\% (i.e. $g^{(2)}(0)$ = 0.12) up to 77~K making the devices suitable for operation using compact Stirling cryo-coolers \cite{Schlehahn2018} and liquid nitrogen, enabling for instance user-friendly QKD applications. The experimentally obtained PEE reaches values up to $(40 \pm 2)\%$ into a numerical aperture (NA) of 0.82 which is in excellent agreement with finite-element method (FEM) simulations yielding PEE = 43\%.

\section{DESIGN, EPITAXIAL GROWTH AND IN-SITU EBL INTEGRATION}

First, we optimized the vertical layer sequence and the CBG geometry for our O-band single-photon sources (SPSs) using the \textsc{JCM}suite FEM solver. The optimization process involved systematically varyies key design parameters, such as the grating period, ring width, etch depth, and mesa diameter to maximize the PEE. Through this parametric study, the solver identifies the configuration that yielded the highest outcoupling efficiency into the collection optics, with an NA of 0.8. We could achieve PEE up to 64\% at a wavelength of 1284~nm and slightly lower values at the central wavelength of the O-band (see Fig.~S1 in the Supplementary Information (SI)). For instance, at the emission wavelength of the investigated QD (1324~nm), the simulation predicts a PEE of 43\%, assuming ideal emitter alignment and a perfectly matched DBR stack. Noteworthy, our CBG design with backside DBR mirror does not yield a significant Purcell effect (see section I of the SI for details on the numerical modelling and results).

Next, we describe the epitaxial growth of the QD-heterostructure and the deterministic nanofabrication of QD-CBG O-band SPSs. The QD-heterostructure with the above-mentioned optimized layer design was fabricated via a two-step epitaxial process using separate MOCVD reactors \cite{Volz2011, Jung2017b, Limame2024b}. Details are provided in the epitaxial growth section of the SI and Fig.~S2. Following thermal deoxidation at 735 °C, a 300-nm-thick GaAs buffer was grown at a high V/III ratio (200). Then, a 32.5-pair GaAs/Al$_{0.90}$Ga$_{0.10}$As DBR was grown to serve as a backside mirror of a later CBG structure. A 192~nm GaAs spacer formed half the optical cavity. QD growth was performed at 500~°C with deposition of a 0.27~nm In$_{0.5}$Ga$_{0.5}$As wetting layer and a 60~s growth interruption for nucleation. The QDs with a density of (1–5 × 10$^{9}$) cm$^{-2}$ were capped by 0.18~nm GaAs and a 4~nm In$_{0.23}$Ga$_{0.77}$As strain-reducing layer to enable larger dots with telecom O-band emission\cite{Srocka2020-ms}. The structure was subsequently annealed at 615~°C to remove residual In, then capped with 190~nm GaAs to complete the cavity.

To integrate single O-band QDs into CBG resonators, the sample surface was first spin-coated at 4000~rpm with the high-resolution electron-beam resist AR-P 6200-13 (CSAR 62, Allresist GmbH). This resulted in a uniform resist layer approximately 400 nm thick, suitable for subsequent EBL and etching steps. To identify the spatial positions of optically active QDs, low-temperature cathodoluminescence (CL) mapping was carried out at 20~K as the first step of the iEBL process. During this step, the electron beam was operated at a low dose of $\approx$ 7.7 mC/cm$^2$ for a pixel size of 500 nm. This dose was chosen as a compromise between achieving sufficient luminescence signal and avoiding resist over- or under-exposure. In this dose regime, the resist still exhibits a positive-tone response, whereby the mapped areas are locally cleared. The subsequent EBL step was performed also at 20~K but at higher doses, where CSAR 62 switches to a negative-tone regime, thereby leaving behind the desired CBG structures aligned to the pre-mapped QDs. Importantly, the exposure strategy incorporated dose modulation to compensate for proximity effects at high doses. After development, the remaining resist served as a hard etch mask for pattern transfer by inductively coupled plasma reactive ion etching (ICP-RIE). 

Using the decribed nanofabrication process, 21 QDs were deterministically integrated into CBG resonators. Eighty percent of the QDs were located within 100 nm of the CBG center (see Section III of the SI and Fig.~S3). Figure~\ref{Fig:CL}(a) presents a CL map of a chosen QD–CBG device overlaid with an SEM image acquired at 20~K. Using Gaussian fits along both the horizontal and vertical axes, we determined the emission maximum of the integrated QD and the center of the mesa, revealing a radial 20~nm offset between the QD position and the CBG mesa center. White and green crosses indicate this offset in the inset of Fig.~\ref{Fig:CL}(a). Additionally, as discussed and shown in the SI, spatial offsets as small as a few nanometers are achieved, demonstrating the high potential and accuracy of iEBL. The high QD–CBG alignment accuracy is crucial for achieving the simulated PEE in the experiment because it enables pronounced light-matter coupling in the cavity QED regime.

Figure~\ref{Fig:CL}(b) shows the CL spectra of the QD-CBG device shown in (a). Multiple emission lines originating from different QDs within the same CBG are visible. The investigated QD, which exhibits a 20~nm offset relative to the mesa center, has an emission wavelength of approximately 1324 nm. The emission features at shorter wavelengths are attributed to other QDs that were unintentionally integrated into the structure due to their spatial proximity to the target QD.

\section{OPTICAL AND QUANTUM OPTICAL INVESTIGATION}

\begin{figure*}[t]  
	\centering
	\includegraphics[width=1\textwidth]{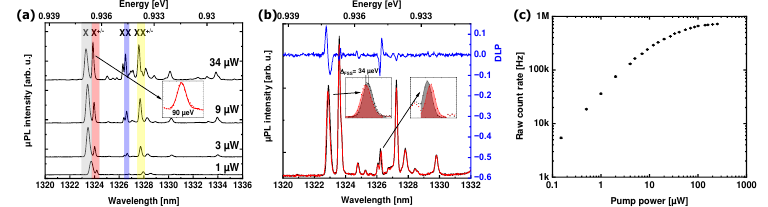}  
	\caption{(a) Waterfall plot of \textmu PL spectra from the QD-CBG device presented in Fig.~\ref{Fig:CL}. The measurements were performed under pulsed excitation at a wavelength of 1240 nm, with increasing excitation powers of 1, 3, 9, 34~\textmu W, at a temperature of 4 K. Four distinct emission lines are visible and are attributed to the neutral exciton (X, black), charged exciton (X$^\pm$, red), biexciton (XX, blue), and charged biexciton (XX$^\pm$, yellow), where the line assignment was performed using excitation and polarization dependent \textmu PL measurements presented in Fig.~S5 panels (a) and (b). Insets show Gaussian fit of the X$^\pm$ line, yielding a linewidth of $(90 \pm 8)$~\textmu eV. (b) \textmu PL spectra of the same QD at an excitation power of 20~\textmu W, measured at two orthogonal polarization angles: 0\textdegree~ (black) and 90\textdegree (red). The insets show emission of the X and XX emission lines, from which an FSS of $(34.2 \pm 0.4)$~\textmu eV is extracted. The degree of linear polarization is plotted in blue, showing the characteristic polarization dependence of the neutral excitonic transitions. (c) Raw count rate of the X$^\pm$ emission as a function of excitation power, as measured with a SNSPD. A maximum count rate of $(0.72~\pm~0.03)$ MHz  was observed at saturation, and was used to determine the PEE of the SPS.}
	\label{Fig:Spectra}
\end{figure*}

In this section, we investigate the optical properties of the device shown in Fig.~\ref{Fig:CL} using \textmu PL spectroscopy. Emission of a pulsed laser with a wavelength of 1240 nm is focused onto the sample through an objective lens (NA = 0.82), producing a beam spot of (1-3)~\textmu m${^2}$. The resulting emission is collected confocally through the same objective and directed to a monochromator with a 150 and 900 lines/mm grating, resulting in a spectral resolution of 150 and 20~\textmu m, respectively. Spectral analysis was performed using a 1D InGaAs line detector. Section IV of the SI and Fig. S4 provide a detailed description and a schematic illustration of the setup. 

Figure~\ref{Fig:Spectra}(a) presents a waterfall plot of \textmu PL emission spectra recorded at excitation powers ranging from 1 to 34~\textmu W and a temperature of 4~K. The spectra reveal distinct spectral lines corresponding to various excitonic complexes: the neutral exciton (X, black), charged excitons or trions (X$^\pm$, red), biexciton (XX, blue), and charged biexciton (XX$^\pm$, yellow). The nature of these states is unambiguously identified via power-dependent and polarization-resolved \textmu PL measurements, as shown in the SI, Fig.~S5. The neutral exciton (X) emission line is centered at 0.9369~eV (1323.32~nm), with a full width at half maximum (FWHM) of $(142 \pm 4)$~\textmu eV. It exhibits a linear input/output (I/O) characteristic and is split by the fine-structure splitting (FSS), which is indicative of its neutral nature. In contrast, the trion (X$^\pm$) at 0.9365~eV 1323.88~nm) lacks an FSS and exhibits a narrower FWHM of $(90 \pm 8)$~\textmu eV. The FSS of the exciton was extracted from the polarization-resolved spectra at 0\textdegree{} and 90\textdegree{}, shown in Fig.~\ref{Fig:Spectra}(b) in black and red, respectively, yielding a value of (34.2~$\pm$~0.4)~\textmu eV. This value is consistent with those previously reported ones for O-band InGaAs/GaAs QDs grown on GaAs substrates \cite{Srocka2020}. The biexciton (XX) line emits at 0.9346~eV (1326.50~nm), while the charged biexciton (XX$^\pm$) is observed at 0.9338~eV (1327.60~nm). Both lines exhibit the characteristic near-quadratic power dependence, with exponents of $(1.31 \pm 0.03)$ for the XX transition and $(1.16 \pm 0.06)$ for the XX$^{+/-}$ transition, respectively. Furthermore, the XX transition shows a linear polarization orthogonal to that of the exciton, as illustrated in Fig.~\ref{Fig:Spectra}(b) and Fig.~S5(b), which further confirms its biexcitonic nature.

In addition to the optical characteristics of the integrated QD, the PEE is a critical figure of merit for applications of quantum light sources. A high PEE directly impacts for instance the achievable communication rates in quantum networks, the generation of photonic cluster states, and the realization of single-photon-based qubits \cite{Couteau2023}. For an NA of 0.8 and an emission wavelength of 1323 nm, the simulated PEE of the fabricated CBG structure is predicted to be 43\%. A schematic of the standard \textmu PL setup used in this study is shown in Fig.~S4 of the SI. The emitted photons from the device are collected with a 0.82~NA objective. A monochromator with a 900~lines/cm grating spectrally resolves the emission, which is then selected using the exit slit and coupled into a fiber beam splitter. The signal is directed onto the superconducting nanowire single-photon detector (SNSPD) detectors for time-resolved as well as autocorrelation measurements. The measured SNSPD count rate of the charged exciton (X$^+$ or X$^-$) transition of (0.72~$\pm$~0.03)~MHz at saturation (see Fig.~\ref{Fig:Spectra}(c)) yields PEE = (22~$\pm$~2)\% using the independently measured setup efficiency of (4.1~$\pm$~0.3)\% and taking the laser repetition rate of 80\,MHz into account. Including also the count rate of the neutral exciton at saturation, the resulting total PEE of QD-CBG device is (40~$\pm$~2)\% and (33~$\pm$~1)\% when taking the non-ideal single-photon purity of (17.2~$\pm$~0.4)\% at saturation into account. The good agreement between the theoretically predicted and experimentally obtained PEE indicates that the QD possesses a high internal quantum efficiency, and demonstrates the optical quality of the InGaAs/GaAs QD and DBR structure grown on a Si substrate. Moreover, it validates the high spatial accuracy of the iEBL integration process used to embed the QD into the photonic structure, as discussed above.

\begin{figure*}[t]  
	\centering
	\includegraphics[width=0.8\textwidth]{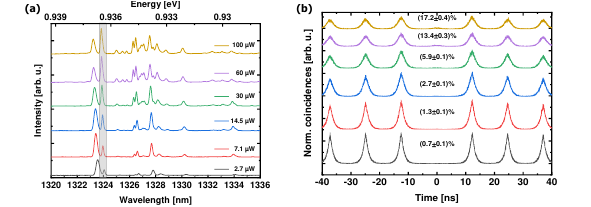}  
	\caption{(a) Waterfall plot of \textmu PL spectra from the investigated structure recorded at excitation powers of 2.7, 7.1, 14.5, 30, 60, and 100~\textmu W. These correspond to the powers at which the second-order photon autocorrelation function $g^{(2)}(\tau)$ shown in panel (b) was measured. All measurements were performed under pulsed excitation at a wavelength of 1240~nm and at a temperature of 4~K. (b) Second-order photon autocorrelation function $g^{(2)}(\tau)$ measured for the charged exciton line at various excitation powers. The corresponding $g^{(2)}(0)$ values are indicated in the legend. The data demonstrate clean single-photon emission behavior, with a $g^{(2)}(0)$ value of $(0.007 \pm 0.001)$ at weak excitation and $(0.172 \pm 0.004)\%$ at saturation power.}
	\label{Fig:LT_g2}
\end{figure*}

 To assess the single-photon purity of the device under study, we performed pulsed second-order correlation measurements using a Hanbury Brown and Twiss (HBT) setup, again under non-resonant pulsed excitation at 1240 nm. Figure~\ref{Fig:LT_g2}(b) shows the resulting $g^{(2)}(\tau)$ correlation histogram, where pronounced antibunching at zero delay is observed for all excitation powers. Figure S6 in the SI shows each of the $g^{(2)}(\tau)$ measurements on a half-logarithmic scale, providing a clearer view of the central $g^{(2)}(0)$ peak. This confirms strong suppression of multi-photon emission. At a low excitation power of 2.7~\textmu W, we determined $g^{(2)}(0) = (0.007 \pm 0.001)$ using the integrated coincidences over 12.5 ns, corresponding to a single-photon purity of $(99.3 \pm 0.1)\%$ \cite{Miyazawa2016}. This remarkably high single-photon purity highlights the emitter's excellent quantum optical quality and the low uncorrelated background contribution from the surrounding material. As the excitation power increases toward saturation (up to 100~\textmu W), we observe an increase in the multi-photon emission probability, with $g^{(2)}(0)$ reaching $(0.172 \pm 0.004)$ at saturation. This behavior is commonly attributed to carrier recapture processes under the applied non-resonant excitation, wherein residual carriers in the wetting layer or barrier material can be captured by the QD, leading to secondary emission events that degrade single-photon purity \cite{Dalgarno2008, Fischbach2017, Holewa2022}. It is known that these effects can be significantly mitigated by using resonant excitation schemes, such as $P$-shell or $S$-shell excitation, two-photon excitation (TPE), or more advanced protocols like SUPER \cite{Huber2015, Karli2022, Hauser2025-gv}. However, performing such experiments is beyond the scope of the present work.

\subsection{TEMPERATURE DEPENDENT PERFORMANCE}

\begin{figure*}  
	\centering
	\includegraphics[width=\textwidth]{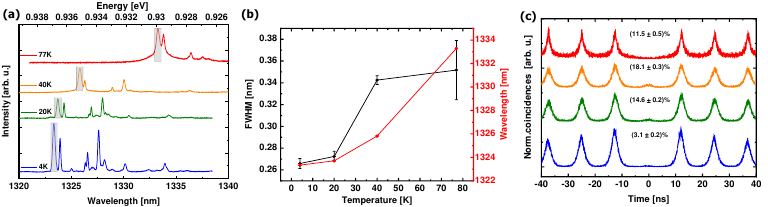}  
	\caption{ (a) Waterfall plot showing the \textmu PL emission from the investigated QD-CBG measured at four different temperatures of 4, 20, 40, 77 K under pulsed excitation. (b) The FWHM (black data points) and emission wavelength (red data points) of the XX\(^{\pm}\) as a function of temperature. (c) Measured second-order photon correlation function $g^{(2)}(\tau)$ at 4, 20, 40, 77~K, demonstrating pure single-photon emission up to 77~K.}
	\label{Fig:TempSer}
\end{figure*}

Finally, we discuss the temperature-dependent optical and quantum optical properties of the QD-CBG. Figure~\ref{Fig:TempSer}(a) presents a waterfall plot of the \textmu  PL spectra under pulsed non-resonant excitation (1240 nm) at similar pump powers (approximately 20 \textmu W) at 4~K (blue), 20~K (green), 40~K (orange), and 77~K (red), respectively. This color scheme is used consistently throughout the remaining figures. The temperatures of 40~K and 77~K are of particular interest for applications: while 40~K can be conveniently reached by compact Stirling cryo-coolers, 77~K corresponds to the boiling point of liquid nitrogen, providing a low-cost and widely available cooling option. As the temperature increases, we generally observe a red shift in the emission due to lowering of the semiconductor bandgap and redistribution of the emission line intensities. Single QD emission lines remain well-pronounced up to 77~K. The neutral exciton (identified by the gray box in Fig.~\ref{Fig:TempSer}(a)) remains the brightest emission line across all temperatures, motivating its selection for additional quantum optical measurements. The progressive spectral broadening observed in Fig.~\ref{Fig:TempSer}(b) (black data points) with increasing temperature can be attributed to enhanced exciton–phonon interactions \cite{Braun2014}, while the concurrent redshift of the emission wavelength (red data points) originates from the temperature-induced lattice expansion, which effectively increases the size of the quantum emitter. These interactions contribute to carrier scattering within the QD and induce dephasing of the excitonic states. As the temperature increases, carriers (electrons and holes) can acquire sufficient thermal energy to escape the quantum confinement potential into the surrounding matrix material. This reduces the population available for radiative recombination within the dot. The combination of thermal activation and phonon-induced dephasing leads to a decrease in the efficiency of single\cite{Denning2019}. 

Time-resolved PL measurements at 4, 20, 40, and 77~K can be found in the SI, section IV. The corresponding second-order autocorrelation measurements are shown in Fig.~\ref{Fig:TempSer}(c). The data display $g^{(2)}(\tau)$ at 1240~nm under 80~MHz pulsed excitation, recorded with a fiber-based HBT setup, as described in detail in the SI. As the temperature increases from 4~K to 40~K, $g^{(2)}(0)$ rises from $(3.1~\pm~0.2)\%$ to $(18.1~\pm~0.3)\%$, which can be attributed to increased uncorrelated background emission and recapture events facilitated by enhanced phonon interaction. These processes lead to phonon-assisted re-excitation of the QD before the system can fully relax to the ground state\cite{Huber2015, Denning2019}. At 77~K, however, the measured $g^{(2)}(0)$ value decreases relative to 40~K, correlating with a reduction in the rise time $t_0$ as discussed in the SI (section IV). A shorter rise time generally indicates that carriers relax more efficiently into the QD ground state, thereby reducing the probability of delayed emission events that would otherwise increase multiphoton probability. At intermediate cryogenic temperatures, such as 40~K, phonon-assisted recapture of carriers from the surrounding wetting layer or barrier states can occur on comparable timescales to radiative recombination. This results in longer $t_0$ and increased $g^{(2)}(0)$ due to re-excitation events\cite{Urayama2001}. 

Interestingly, at higher temperatures around 77~K the increased phonon population accelerates carrier thermalization and relaxation into the lowest QD states. This suppresses effectively long-lived recapture channels and leads to faster buildup of the emission signal. This behavior illustrates the delicate balance between phonon-assisted recapture, which reduces the single-photon purity, and phonon-mediated relaxation, which, under certain conditions, can enhance the temporal fidelity of exciton initialization.

\section{SUMMARY AND CONCLUSION}

In conclusion, we have demonstrated the epitaxial growth of high-quality InGaAs/GaAs quantum dots on silicon, enabling single-photon emission in the telecommunication O-band. In the developed growth protocol, the GaP buffer layer and optimized interface engineering ensured optically active, dislocation-suppressed QDs on Si, while the InGaAs strain-reducing layer provided emission tuning into the telecom band.
Deterministic in-situ electron-beam lithography allowed accurate placement of individual emitters into circular Bragg grating resonators, yielding high photon extraction efficiencies and a strong agreement between simulated and measured PEE values of 43\% and (40 $\pm$ 1)\%, respectively. The devices exhibit excellent optical and quantum-optical performance, including high single-photon purity under non-resonant excitation and robust operation at elevated temperatures, as evidenced by $g^{(2)}(0)$ = (14.6 $\pm$ 0.2)\% at 40 K and (11.5 $\pm$ 0.5)\% at 77~K. These results highlight the practicality and reliability of this integration strategy and position InGaAs/GaAs QDs on Si as a competitive and scalable alternative to GaAs-based platforms. Overall, the demonstrated approach marks an important step toward high-performance, large-scale, and cost-effective quantum photonic circuits compatible with silicon photonics.

\section*{ACKNOWLEDGMENTS}

The research leading to these results received funding from the BMFTR via projects 16KISQ014 and 16KISQ087K, from the German Research Foundation via projects Re2974/23-1, INST 131/795-1 320 FUGG  and from Berlin Quantum. The authors would also thank Kathrin Schatke, Praphat Sonka, Heike Oppermann, Stefan Bock for their technical support. The authors further acknowledge Martin Von Helversen, Lucas Rickert, and Daniel Vajner for their invaluable support and engaging scientific discussions.

\section*{DATA AVAILABILITY STATEMENT}

The data that support the findings of this study are available from the corresponding author upon reasonable request.

\section*{Author Declarations}
The authors have no conflicts to disclose.

\section*{REFERENCES}
	\bibliographystyle{unsrt}
	\bibliography{References}

@PREAMBLE{
 "\providecommand{\noopsort}[1]{}" 
 # "\providecommand{\singleletter}[1]{#1}%" 
}

@article{Arakawa2020,
  title = {Progress in quantum-dot single photon sources for quantum information technologies: A broad spectrum overview},
  volume = {7},
  ISSN = {1931-9401},
  url = {http://dx.doi.org/10.1063/5.0010193},
  DOI = {10.1063/5.0010193},
  number = {2},
  journal = {Applied Physics Reviews},
  publisher = {AIP Publishing},
author = {Arakawa, Y. and Holmes, M. J.},
  year = {2020},
  month = jun 
}

@article{Schimpf2021,
  title = {Quantum cryptography with highly entangled photons from semiconductor quantum dots},
  volume = {7},
  ISSN = {2375-2548},
  url = {http://dx.doi.org/10.1126/sciadv.abe8905},
  DOI = {10.1126/sciadv.abe8905},
  number = {16},
  journal = {Science Advances},
  publisher = {American Association for the Advancement of Science (AAAS)},
 author = {Schimpf, C. and Reindl, M. and Huber, D. and Lehner, B. and Covre Da Silva, S. F. and Manna, S. and Vyvlecka, M. and Walther, P. and Rastelli, A.},
  year = {2021},
  month = apr 
}

@inbook{Reitzenstein2025,
  title = {Generation of indistinguishable photons with semiconductor quantum dots},
  ISBN = {9780323958196},
  url = {http://dx.doi.org/10.1016/B978-0-323-96027-4.00022-X},
  DOI = {10.1016/b978-0-323-96027-4.00022-x},
  booktitle = {Comprehensive Semiconductor Science and Technology},
  publisher = {Elsevier},
  author = {Reitzenstein,  S.},
  year = {2025},
  pages = {689–732}
}

@article{Singh2024,
  author       = {Singh},
  title        = {The Race for Quantum Technology in Asia},
  journal      = {Defence \& Diplomacy},
  year         = {2024},
  volume       = {13},
  number       = {3},
  pages        = {77--88},
  month        = jun,
  url          = {https://journals.capsindia.org/index.php/dnd/article/view/25}
}

@Inbook{Lele2021,
author="Lele, A.",
title="Quantum (Arms) Race",
bookTitle="Quantum Technologies and Military Strategy",
year="2021",
publisher="Springer International Publishing",
address="Cham",
pages="145--172",
isbn="978-3-030-72721-5",
doi="10.1007/978-3-030-72721-5_9",
url="https://doi.org/10.1007/978-3-030-72721-5_9"
}

@techreport{Krause2024,
  author       = {Krause, J.},
  title        = {The Quantum Race: U.S.-Chinese Competition for Leadership in Quantum Technologies},
  institution  = {University of California Institute on Global Conflict and Cooperation, UC San Diego},
  year         = {2024},
  url          = {https://escholarship.org/uc/item/4fd2j1k9},
  note         = {Accessed: 21 July 2025}
}

@ARTICLE{Shields2007-wl,
  title     = "Semiconductor quantum light sources",
  author    = "Shields, A. J",
  journal   = "Nature Photonics",
  publisher = "Springer Science and Business Media LLC",
  volume    =  1,
  number    =  4,
  pages     = "215--223",
  month     =  apr,
  year      =  2007,
  language  = "en"
}

@ARTICLE{Senellart2017-mc,
  title     = "High-performance semiconductor quantum-dot single-photon sources",
  author    = "Senellart, P. and Solomon, G. and White, A.",
  journal   = "Nature Nanotechnology",
  publisher = "Springer Science and Business Media LLC",
  volume    =  12,
  number    =  11,
  pages     = "1026--1039",
  month     =  nov,
  year      =  2017,
  language  = "en"
}

@ARTICLE{Srocka2020-ms,
  title     = "Deterministically fabricated quantum dot single-photon source
               emitting indistinguishable photons in the telecom {O}-band",
author = {Srocka, N. and Mrowiński, P. and Große, J. and von Helversen, M. and Heindel, T. and Rodt, S. and Reitzenstein, S.},
  journal   = "Applied Physics Letters",
  publisher = "AIP Publishing",
  volume    =  116,
  number    =  23,
  pages     = "231104",
  month     =  jun,
  year      =  2020,
  language  = "en"
}

@ARTICLE{Hauser2025-gv,
  title        = "Deterministic and highly indistinguishable single photons in
                  the telecom {C}-band",
  author = {Hauser, N. and Bayerbach, M. and Kaupp, J. and Reum, Y. and Peniakov, G. and Michl, J. and Kamp, M. and Huber-Loyola, T. and Pfenning, A. T. and Höfling, S. and Barz, S.},
  year         =  2025,
  primaryClass = "quant-ph",
    journal   = "ArXiv",
doi = {10.48550/ARXIV.2505.09695},
  url = {https://arxiv.org/abs/2505.09695},
  eprint       = "2505.09695"
}

@ARTICLE{Alqedra2025-fh,
  title    = "Entangled photon pair generation in the telecom {O}-band from
              nanowire quantum dots",
  author = {Alqedra, M. K. and Huang, C.-T. and Yeung, E. and Chang, W.-H. and Haffouz, S. and Poole, P. J. and Dalacu, D. and Elshaari, A. W. and Zwiller, V.},
  journal  = "Nano Letters",
  volume   =  25,
  number   =  26,
  pages    = "10321--10327",
  month    =  jul,
  year     =  2025,
  language = "en"
}

@ARTICLE{Rau2014-pc,
  title     = "Free space quantum key distribution over 500 meters using
               electrically driven quantum dot single-photon sources---a proof
               of principle experiment",
  author = {Rau, M. and Heindel, T. and Unsleber, S. and Braun, T. and Fischer, J. and Frick, S. and Nauerth, S. and Schneider, C. and Vest, G. and Reitzenstein, S. and Kamp, M. and Forchel, A. and Höfling, S. and Weinfurter, H.},
  journal   = "New Journal of Physics",
  publisher = "IOP Publishing",
  volume    =  16,
  number    =  4,
  pages     = "043003",
  month     =  apr,
  year      =  2014,
  copyright = "http://creativecommons.org/licenses/by/3.0/"
}

@ARTICLE{Zhang2025-vo,
  title     = "Experimental single-photon quantum key distribution surpassing
               the fundamental weak coherent-state rate limit",
  author = {Zhang, Y. and Ding, X. and Li, Y. and Zhang, L. and Guo, Y.-P. and Wang, G.-Q. and Ning, Z. and Xu, M.-C. and Liu, R.-Z. and Zhao, J.-Y. and Zou, G.-Y. and Wang, H. and Cao, Y. and He, Y.-M. and Peng, C.-Z. and Huo, Y.-H. and Liao, S.-K. and Lu, C.-Y. and Xu, F. and Pan, J.-W.},
  journal   = "Physical Review Letters",
  publisher = "American Physical Society (APS)",
  volume    =  134,
  number    =  21,
  pages     = "210801",
  month     =  may,
  year      =  2025,
  copyright = "https://link.aps.org/licenses/aps-default-license",
  language  = "en"
}

@ARTICLE{Maring2024-ia,
  title     = "A versatile single-photon-based quantum computing platform",
  author = {Maring, N. and Fyrillas, A. and Pont, M. and Ivanov, E. and Stepanov, P. and Margaria, N. and Hease, W. and Pishchagin, A. and Lemaître, A. and Sagnes, I. and Au, T. H. and Boissier, S. and Bertasi, E. and Baert, A. and Valdivia, M. and Billard, M. and Acar, O. and Brieussel, A. and Mezher, R. and Wein, S. C. and Salavrakos, A. and Sinnott, P. and Fioretto, D. A. and Emeriau, P.-E. and Belabas, N. and Mansfield, S. and Senellart, P. and Senellart, J. and Somaschi, N.},
  journal   = "Nature Photonics",
  publisher = "Springer Science and Business Media LLC",
  volume    =  18,
  number    =  6,
  pages     = "603--609",
  month     =  jun,
  year      =  2024,
  copyright = "https://creativecommons.org/licenses/by/4.0",
  language  = "en"
}

@ARTICLE{Couteau2023-sl,
  title     = "Applications of single photons to quantum communication and
               computing",
  author    = "Couteau, C. and Barz, S. and Durt, T. and Gerrits, T. and Huwer, J. and Prevedel, R. and Rarity, J. and Shields, A. and Weihs, G.",
  journal   = "Nature Reviews Physics",
  publisher = "Springer Science and Business Media LLC",
  volume    =  5,
  number    =  6,
  pages     = "326--338",
  month     =  may,
  year      =  2023,
  language  = "en"
}

@article{Katsumi2019,
  title     = {Quantum-dot single-photon source on a CMOS silicon photonic chip integrated using transfer printing},
  volume    = {4},
  ISSN      = {2378-0967},
  url       = {http://dx.doi.org/10.1063/1.5087263},
  DOI       = {10.1063/1.5087263},
  number    = {3},
  journal   = {APL Photonics},
  publisher = {AIP Publishing},
  author    = {Katsumi, R. and Ota, Y. and Osada, A. and Yamaguchi, T. and Tajiri, T. and Kakuda, M. and Iwamoto, S. and Akiyama, H. and Arakawa, Y.},
  year      = {2019},
  month     = mar
}

@article{Vijayan2024,
  title     = {Growth of telecom {C}-band {In(Ga)As} quantum dots for silicon quantum photonics},
  volume    = {4},
  ISSN      = {2633-4356},
  url       = {http://dx.doi.org/10.1088/2633-4356/ad2522},
  DOI       = {10.1088/2633-4356/ad2522},
  number    = {1},
  journal   = {Materials for Quantum Technology},
  publisher = {IOP Publishing},
  author    = {Vijayan, P. and Joos, R. and Werner, M. and Hirlinger-Alexander, J. and Seibold, M. and Vollmer, S. and Sittig, R. and Bauer, S. and Braun, F. and Portalupi, S. L. and Jetter, M. and Michler, P.},
  year      = {2024},
  month     = feb,
  pages     = {016301}
}

@article{Volz2011,
  title     = {{GaP}-nucleation on exact {Si} (001) substrates for {III/V} device integration},
  volume    = {315},
  ISSN      = {0022-0248},
  url       = {http://dx.doi.org/10.1016/j.jcrysgro.2010.10.036},
  DOI       = {10.1016/j.jcrysgro.2010.10.036},
  number    = {1},
  journal   = {Journal of Crystal Growth},
  publisher = {Elsevier BV},
  author    = {Volz, K. and Beyer, A. and Witte, W. and Ohlmann, J. and Németh, I. and Kunert, B. and Stolz, W.},
  year      = {2011},
  month     = jan,
  pages     = {37--47}
}

@article{Beyer2013,
  title     = {Atomic structure of (110) anti-phase boundaries in {GaP} on {Si}(001)},
  volume    = {103},
  ISSN      = {1077-3118},
  url       = {http://dx.doi.org/10.1063/1.4815985},
  DOI       = {10.1063/1.4815985},
  number    = {3},
  journal   = {Applied Physics Letters},
  publisher = {AIP Publishing},
  author    = {Beyer, A. and Haas, B. and Gries, K. I. and Werner, K. and Luysberg, M. and Stolz, W. and Volz, K.},
  year      = {2013},
  month     = jul
}

@article{Beyer2011,
  title     = {Influence of crystal polarity on crystal defects in {GaP} grown on exact {Si} (001)},
  volume    = {109},
  ISSN      = {1089-7550},
  url       = {http://dx.doi.org/10.1063/1.3567910},
  DOI       = {10.1063/1.3567910},
  number    = {8},
  journal   = {Journal of Applied Physics},
  publisher = {AIP Publishing},
  author    = {Beyer, A. and Németh, I. and Liebich, S. and Ohlmann, J. and Stolz, W. and Volz, K.},
  year      = {2011},
  month     = apr
}

@article{Nmeth2008,
  title     = {Heteroepitaxy of GaP on Si: Correlation of morphology, anti-phase-domain structure and MOVPE growth conditions},
  volume    = {310},
  ISSN      = {0022-0248},
  url       = {http://dx.doi.org/10.1016/j.jcrysgro.2007.11.127},
  DOI       = {10.1016/j.jcrysgro.2007.11.127},
  number    = {7--9},
  journal   = {Journal of Crystal Growth},
  publisher = {Elsevier BV},
  author    = {Németh, I. and Kunert, B. and Stolz, W. and Volz, K.},
  year      = {2008},
  month     = apr,
  pages     = {1595--1601}
}

@article{Huang2014,
  title     = {{InGaAs/GaAs} quantum well lasers grown on exact {GaP/Si} (001)},
  volume    = {50},
  ISSN      = {1350-911X},
  url       = {http://dx.doi.org/10.1049/el.2014.2077},
  DOI       = {10.1049/el.2014.2077},
  number    = {17},
  journal   = {Electronics Letters},
  publisher = {Institution of Engineering and Technology (IET)},
  author    = {Huang, X. and Song, Y. and Masuda, T. and Jung, D. and Lee, M.},
  year      = {2014},
  month     = aug,
  pages     = {1226--1227}
}

@article{Shang2021,
  title     = {High-temperature reliable quantum-dot lasers on {Si} with misfit and threading dislocation filters},
  volume    = {8},
  ISSN      = {2334-2536},
  url       = {http://dx.doi.org/10.1364/OPTICA.423360},
  DOI       = {10.1364/optica.423360},
  number    = {5},
  journal   = {Optica},
  publisher = {Optica Publishing Group},
  author    = {Shang, C. and Hughes, E. and Wan, Y. and Dumont, M. and Koscica, R. and Selvidge, J. and Herrick, R. and Gossard, A. C. and Mukherjee, K. and Bowers, J. E.},
  year      = {2021},
  month     = may,
  pages     = {749}
}

@article{Jung2017,
  title     = {Highly Reliable Low-Threshold {InAs} Quantum Dot Lasers on On-Axis (001) {Si} with 87% Injection Efficiency},
  volume    = {5},
  ISSN      = {2330-4022},
  url       = {http://dx.doi.org/10.1021/acsphotonics.7b01387},
  DOI       = {10.1021/acsphotonics.7b01387},
  number    = {3},
  journal   = {ACS Photonics},
  publisher = {American Chemical Society (ACS)},
  author    = {Jung, D. and Zhang, Z. and Norman, J. and Herrick, R. and Kennedy, M. J. and Patel, P. and Turnlund, K. and Jan, C. and Wan, Y. and Gossard, A. C. and Bowers, J. E.},
  year      = {2017},
  month     = dec,
  pages     = {1094--1100}
}

@article{Limame2024b,
  title     = {High-quality single {InGaAs/GaAs} quantum dot growth on a silicon substrate for quantum photonic applications},
  volume    = {2},
  ISSN      = {2837-6714},
  url       = {http://dx.doi.org/10.1364/OPTICAQ.510829},
  DOI       = {10.1364/opticaq.510829},
  number    = {2},
  journal   = {Optica Quantum},
  publisher = {Optica Publishing Group},
  author    = {Limame, I. and Ludewig, P. and Shih, C.-W. and Hohn, M. and Palekar, C. C. and Stolz, W. and Reitzenstein, S.},
  year      = {2024},
  month     = apr,
  pages     = {117}
}

@article{Srocka2020,
  title     = {Deterministically fabricated strain-tunable quantum dot single-photon sources emitting in the telecom  {O}-band},
  volume    = {117},
  ISSN      = {1077-3118},
  url       = {http://dx.doi.org/10.1063/5.0030991},
  DOI       = {10.1063/5.0030991},
  number    = {22},
  journal   = {Applied Physics Letters},
  publisher = {AIP Publishing},
  author    = {Srocka, N. and Mrowiński, P. and Große, J. and Schmidt, M. and Rodt, S. and Reitzenstein, S.},
  year      = {2020},
  month     = nov
}

@article{Donges2022,
  title     = {Machine learning enhanced in situ electron beam lithography of photonic nanostructures},
  volume    = {14},
  ISSN      = {2040-3372},
  url       = {http://dx.doi.org/10.1039/D2NR03696G},
  DOI       = {10.1039/d2nr03696g},
  number    = {39},
  journal   = {Nanoscale},
  publisher = {Royal Society of Chemistry (RSC)},
  author    = {Donges, J. and Schlischka, M. and Shih, C.-W. and Pengerla, M. and Limame, I. and Schall, J. and Bremer, L. and Rodt, S. and Reitzenstein, S.},
  year      = {2022},
  pages     = {14529--14536}
}

@article{Engel2023,
  title     = {Purcell enhanced single-photon emission from a quantum dot coupled to a truncated Gaussian microcavity},
  volume    = {122},
  ISSN      = {1077-3118},
  url       = {http://dx.doi.org/10.1063/5.0128631},
  DOI       = {10.1063/5.0128631},
  number    = {4},
  journal   = {Applied Physics Letters},
  publisher = {AIP Publishing},
  author    = {Engel, L. and Kolatschek, S. and Herzog, T. and Vollmer, S. and Jetter, M. and Portalupi, S. L. and Michler, P.},
  year      = {2023},
  month     = jan
}

@article{Schlehahn2018,
  title     = {A stand-alone fiber-coupled single-photon source},
  volume    = {8},
  ISSN      = {2045-2322},
  url       = {http://dx.doi.org/10.1038/s41598-017-19049-4},
  DOI       = {10.1038/s41598-017-19049-4},
  number    = {1},
  journal   = {Scientific Reports},
  publisher = {Springer Science and Business Media LLC},
  author    = {Schlehahn, A. and Fischbach, S. and Schmidt, R. and Kaganskiy, A. and Strittmatter, A. and Rodt, S. and Heindel, T. and Reitzenstein, S.},
  year      = {2018},
  month     = jan
}

@article{Jung2017b,
  title     = {Low threading dislocation density {GaAs} growth on on-axis {GaP/Si} (001)},
  volume    = {122},
  ISSN      = {1089-7550},
  url       = {http://dx.doi.org/10.1063/1.5001360},
  DOI       = {10.1063/1.5001360},
  number    = {22},
  journal   = {Journal of Applied Physics},
  publisher = {AIP Publishing},
  author    = {Jung, D. and Callahan, P. G. and Shin, B. and Mukherjee, K. and Gossard, A. C. and Bowers, J. E.},
  year      = {2017},
  month     = dec
}

@manual{jcmwave2025,
  title  = {{JCMsuite} – Simulation Software for Nano-Optics, Photonics and Electromagnetics},
  author = {{JCMwave GmbH}},
  year   = {2025},
  note   = {Accessed: August 5, 2025},
  url    = {https://jcmwave.com/docs/}
}

@article{Rickert2019,
  title     = {Optimized designs for telecom-wavelength quantum light sources based on hybrid circular Bragg gratings},
  volume    = {27},
  ISSN      = {1094-4087},
  url       = {http://dx.doi.org/10.1364/OE.27.036824},
  DOI       = {10.1364/oe.27.036824},
  number    = {25},
  journal   = {Optics Express},
  publisher = {Optica Publishing Group},
  author    = {Rickert, L. and Kupko, T. and Rodt, S. and Reitzenstein, S. and Heindel, T.},
  year      = {2019},
  month     = dec,
  pages     = {36824}
}

@article{Madigawa2024,
  title     = {Assessing the Alignment Accuracy of State-of-the-Art Deterministic Fabrication Methods for Single Quantum Dot Devices},
  volume    = {11},
  ISSN      = {2330-4022},
  url       = {http://dx.doi.org/10.1021/acsphotonics.3c01368},
  DOI       = {10.1021/acsphotonics.3c01368},
  number    = {3},
  journal   = {ACS Photonics},
  publisher = {American Chemical Society (ACS)},
  author    = {Madigawa, A. A. and Donges, J. N. and Gaál, B. and Li, S. and Jacobsen, M. A. and Liu, H. and Dai, D. and Su, X. and Shang, X. and Ni, H. and Schall, J. and Rodt, S. and Niu, Z. and Gregersen, N. and Reitzenstein, S. and Munkhbat, B.},
  year      = {2024},
  month     = feb,
  pages     = {1012--1023}
}

@article{Couteau2023,
  title     = {Applications of single photons to quantum communication and computing},
  volume    = {5},
  ISSN      = {2522-5820},
  url       = {http://dx.doi.org/10.1038/s42254-023-00583-2},
  DOI       = {10.1038/s42254-023-00583-2},
  number    = {6},
  journal   = {Nature Reviews Physics},
  publisher = {Springer Science and Business Media LLC},
  author    = {Couteau, C. and Barz, S. and Durt, T. and Gerrits, T. and Huwer, J. and Prevedel, R. and Rarity, J. and Shields, A. and Weihs, G.},
  year      = {2023},
  month     = may,
  pages     = {326--338}
}

@article{Groe2021,
  title     = {Quantum efficiency and oscillator strength of {InGaAs} quantum dots for single-photon sources emitting in the telecommunication {O}-band},
  volume    = {119},
  ISSN      = {1077-3118},
  url       = {http://dx.doi.org/10.1063/5.0059458},
  DOI       = {10.1063/5.0059458},
  number    = {6},
  journal   = {Applied Physics Letters},
  publisher = {AIP Publishing},
  author    = {Große, J. and Mrowiński, P. and Srocka, N. and Reitzenstein, S.},
  year      = {2021},
  month     = aug
}

@article{Maisch2024,
  title     = {Investigation of Purcell enhancement of quantum dots emitting in the telecom {O}-band with an open fiber cavity},
  volume    = {110},
  ISSN      = {2469-9969},
  url       = {http://dx.doi.org/10.1103/PhysRevB.110.165301},
  DOI       = {10.1103/physrevb.110.165301},
  number    = {16},
  journal   = {Physical Review B},
  publisher = {American Physical Society (APS)},
  author    = {Maisch, J. and Grammel, J. and Tran, N. and Jetter, M. and Portalupi, S. L. and Hunger, D. and Michler, P.},
  year      = {2024},
  month     = oct
}

@article{Yu2023,
  title     = {Telecom-band quantum dot technologies for long-distance quantum networks},
  volume    = {18},
  ISSN      = {1748-3395},
  url       = {http://dx.doi.org/10.1038/s41565-023-01528-7},
  DOI       = {10.1038/s41565-023-01528-7},
  number    = {12},
  journal   = {Nature Nanotechnology},
  publisher = {Springer Science and Business Media LLC},
  author    = {Yu, Y. and Liu, S. and Lee, C.-M. and Michler, P. and Reitzenstein, S. and Srinivasan, K. and Waks, E. and Liu, J.},
  year      = {2023},
  month     = dec,
  pages     = {1389--1400}
}

@article{Holewa2022,
  title     = {Bright Quantum Dot Single-Photon Emitters at Telecom Bands Heterogeneously Integrated on {Si}},
  volume    = {9},
  ISSN      = {2330-4022},
  url       = {http://dx.doi.org/10.1021/acsphotonics.2c00027},
  DOI       = {10.1021/acsphotonics.2c00027},
  number    = {7},
  journal   = {ACS Photonics},
  publisher = {American Chemical Society (ACS)},
  author    = {Holewa, P. and Sakanas, A. and G\"{u}r, U. M. and Mrowiński, P. and Huck, A. and Wang, B.-Y. and Musiał, A. and Yvind, K. and Gregersen, N. and Syperek, M. and Semenova, E.},
  year      = {2022},
  month     = jun,
  pages     = {2273--2279}
}

@article{Dalgarno2008,
  title     = {Hole recapture limited single photon generation from a single n-type charge-tunable quantum dot},
  volume    = {92},
  ISSN      = {1077-3118},
  url       = {http://dx.doi.org/10.1063/1.2924315},
  DOI       = {10.1063/1.2924315},
  number    = {19},
  journal   = {Applied Physics Letters},
  publisher = {AIP Publishing},
  author    = {Dalgarno, P. A. and McFarlane, J. and Brunner, D. and Lambert, R. W. and Gerardot, B. D. and Warburton, R. J. and Karrai, K. and Badolato, A. and Petroff, P. M.},
  year      = {2008},
  month     = may
}

@article{Fischbach2017,
  title     = {Single Quantum Dot with Microlens and 3D-Printed Micro-objective as Integrated Bright Single-Photon Source},
  volume    = {4},
  ISSN      = {2330-4022},
  url       = {http://dx.doi.org/10.1021/acsphotonics.7b00253},
  DOI       = {10.1021/acsphotonics.7b00253},
  number    = {6},
  journal   = {ACS Photonics},
  publisher = {American Chemical Society (ACS)},
  author    = {Fischbach, S. and Schlehahn, A. and Thoma, A. and Srocka, N. and Gissibl, T. and Ristok, S. and Thiele, S. and Kaganskiy, A. and Strittmatter, A. and Heindel, T. and Rodt, S. and Herkommer, A. and Giessen, H. and Reitzenstein, S.},
  year      = {2017},
  month     = jun,
  pages     = {1327--1332}
}

@article{Karli2022,
  title     = {SUPER Scheme in Action: Experimental Demonstration of Red-Detuned Excitation of a Quantum Emitter},
  volume    = {22},
  ISSN      = {1530-6992},
  url       = {http://dx.doi.org/10.1021/acs.nanolett.2c01783},
  DOI       = {10.1021/acs.nanolett.2c01783},
  number    = {16},
  journal   = {Nano Letters},
  publisher = {American Chemical Society (ACS)},
  author    = {Karli, Y. and Kappe, F. and Remesh, V. and Bracht, T. K. and M\"{u}nzberg, J. and Covre da Silva, S. and Seidelmann, T. and Axt, V. M. and Rastelli, A. and Reiter, D. E. and Weihs, G.},
  year      = {2022},
  month     = jul,
  pages     = {6567--6572}
}

@article{Huber2015,
  title     = {Optimal excitation conditions for indistinguishable photons from quantum dots},
  volume    = {17},
  ISSN      = {1367-2630},
  url       = {http://dx.doi.org/10.1088/1367-2630/17/12/123025},
  DOI       = {10.1088/1367-2630/17/12/123025},
  number    = {12},
  journal   = {New Journal of Physics},
  publisher = {IOP Publishing},
  author    = {Huber, T. and Predojević, A. and F\"{o}ger, D. and Solomon, G. and Weihs, G.},
  year      = {2015},
  month     = dec,
  pages     = {123025}
}

@article{Rodt2021,
  title     = {High-performance deterministic in situ electron-beam lithography enabled by cathodoluminescence spectroscopy},
  volume    = {2},
  ISSN      = {2632-959X},
  url       = {http://dx.doi.org/10.1088/2632-959X/abed3c},
  DOI       = {10.1088/2632-959x/abed3c},
  number    = {1},
  journal   = {Nano Express},
  publisher = {IOP Publishing},
  author    = {Rodt, S. and Reitzenstein, S.},
  year      = {2021},
  month     = mar,
  pages     = {014007}
}

@article{Bogdanov2016,
  title     = {Material platforms for integrated quantum photonics},
  volume    = {7},
  ISSN      = {2159-3930},
  url       = {http://dx.doi.org/10.1364/OME.7.000111},
  DOI       = {10.1364/ome.7.000111},
  number    = {1},
  journal   = {Optical Materials Express},
  publisher = {Optica Publishing Group},
  author    = {Bogdanov, S. and Shalaginov, M. Y. and Boltasseva, A. and Shalaev, V. M.},
  year      = {2016},
  month     = dec,
  pages     = {111}
}

@article{Schweickert2018,
  title     = {On-demand generation of background-free single photons from a solid-state source},
  volume    = {112},
  ISSN      = {1077-3118},
  url       = {http://dx.doi.org/10.1063/1.5020038},
  DOI       = {10.1063/1.5020038},
  number    = {9},
  journal   = {Applied Physics Letters},
  publisher = {AIP Publishing},
  author    = {Schweickert, L. and J\"{o}ns, K. D. and Zeuner, K. D. and Covre da Silva, S. F. and Huang, H. and Lettner, T. and Reindl, M. and Zichi, J. and Trotta, R. and Rastelli, A. and Zwiller, V.},
  year      = {2018},
  month     = feb
}

@article{Crawford2021,
  title     = {Quantum Sensing for Energy Applications: Review and Perspective},
  volume    = {4},
  ISSN      = {2511-9044},
  url       = {http://dx.doi.org/10.1002/qute.202100049},
  DOI       = {10.1002/qute.202100049},
  number    = {8},
  journal   = {Advanced Quantum Technologies},
  publisher = {Wiley},
  author    = {Crawford, S. E. and Shugayev, R. A. and Paudel, H. P. and Lu, P. and Syamlal, M. and Ohodnicki, P. R. and Chorpening, B. and Gentry, R. and Duan, Y.},
  year      = {2021},
  month     = jun
}

@article{Braun2014,
  title     = {Temperature dependency of the emission properties from positioned {In(Ga)As/GaAs} quantum dots},
  volume    = {4},
  ISSN      = {2158-3226},
  url       = {http://dx.doi.org/10.1063/1.4896284},
  DOI       = {10.1063/1.4896284},
  number    = {9},
  journal   = {AIP Advances},
  publisher = {AIP Publishing},
  author    = {Braun, T. and Schneider, C. and Maier, S. and Igusa, R. and Iwamoto, S. and Forchel, A. and H\"{o}fling, S. and Arakawa, Y. and Kamp, M.},
  year      = {2014},
  month     = sep
}

@article{Denning2019,
  title     = {Phonon effects in quantum dot single-photon sources},
  volume    = {10},
  ISSN      = {2159-3930},
  url       = {http://dx.doi.org/10.1364/OME.380601},
  DOI       = {10.1364/ome.380601},
  number    = {1},
  journal   = {Optical Materials Express},
  publisher = {Optica Publishing Group},
  author    = {Denning, E. V. and Iles-Smith, J. and Gregersen, N. and Mork, J.},
  year      = {2019},
  month     = dec,
  pages     = {222}
}

@article{Urayama2001,
  title     = {Observation of Phonon Bottleneck in Quantum Dot Electronic Relaxation},
  volume    = {86},
  ISSN      = {1079-7114},
  url       = {http://dx.doi.org/10.1103/PhysRevLett.86.4930},
  DOI       = {10.1103/physrevlett.86.4930},
  number    = {21},
  journal   = {Physical Review Letters},
  publisher = {American Physical Society (APS)},
  author    = {Urayama, J. and Norris, T. B. and Singh, J. and Bhattacharya, P.},
  year      = {2001},
  month     = may,
  pages     = {4930--4933}
}

@article{Miyazawa2016,
  title     = {Single-photon emission at 1.5 $\mu$m from an InAs/InP quantum dot with highly suppressed multi-photon emission probabilities},
  volume    = {109},
  ISSN      = {1077-3118},
  url       = {http://dx.doi.org/10.1063/1.4961888},
  DOI       = {10.1063/1.4961888},
  number    = {13},
  journal   = {Applied Physics Letters},
  publisher = {AIP Publishing},
  author    = {Miyazawa, T. and Takemoto, K. and Nambu, Y. and Miki, S. and Yamashita, T. and Terai, H. and Fujiwara, M. and Sasaki, M. and Sakuma, Y. and Takatsu, M. and Yamamoto, T. and Arakawa, Y.},
  year      = {2016},
  month     = sep
}

@article{Heitz2001,
  title = {Existence of a phonon bottleneck for excitons in quantum dots},
  volume = {64},
  ISSN = {1095-3795},
  url = {http://dx.doi.org/10.1103/PhysRevB.64.241305},
  DOI = {10.1103/physrevb.64.241305},
  number = {24},
  journal = {Physical Review B},
  publisher = {American Physical Society (APS)},
  author = {Heitz,  R. and Born,  H. and Guffarth,  F. and Stier,  O. and Schliwa,  A. and Hoffmann,  A. and Bimberg,  D.},
  year = {2001},
  month = nov 
}

\end{document}


\preprint{AIP/123-QED}

\title{Direct Epitaxial Growth and Deterministic Device Integration of high-quality Telecom O-Band InGaAs Quantum Dots on Silicon Substrate: Supplementary Information}
	\author{I. Limame}
	\altaffiliation{imad.limame@tu-berlin.de}
	\email{stephan.reitzenstein@physik.tu-berlin.de}
    \affiliation{Institute of Physics and Astronomy, Technical University of Berlin, Hardenbergstraße 36, D-10623 Berlin, Germany}
    \author{P. Ludewig}
    \affiliation{NAsP III/V GmbH, Hans-Meerwein-Straße 6, D-35032 Marburg, Germany}
    \author{A. Koulas-Simos}
	\author{C. C. Palekar}
    \author{J. Donges} 
    \author{C.-W. Shih} 
	\author{K. Gaur}
	\author{S. Tripathi}
        \author{S. Rodt}
   \affiliation{Institute of Physics and Astronomy, Technical University of Berlin, Hardenbergstraße 36, D-10623 Berlin, Germany}
    \author{W. Stolz}
	\affiliation{NAsP III/V GmbH, Hans-Meerwein-Straße 6, D-35032 Marburg, Germany}
    \author{K. Volz}
	\affiliation{mar.quest | Marburg Center for Quantum Materials and Sustainable Technologies, Philipps University Marburg, 35032 Marburg, Germany}
    \affiliation{Department of Physics, Philipps University Marburg, Hans Meerwein Str. 6, 35032 Marburg, Germany}
	\author{S. Reitzenstein}
	\altaffiliation{stephan.reitzenstein@physik.tu-berlin.de}
	\affiliation{Institute of Physics and Astronomy, Technical University of Berlin, Hardenbergstraße 36, D-10623 Berlin, Germany}
\maketitle
\date{\today}

\section{NUMERICAL MODELLING}

\begin{figure*}[t]  
	\centering
	\includegraphics[width=0.8\textwidth]{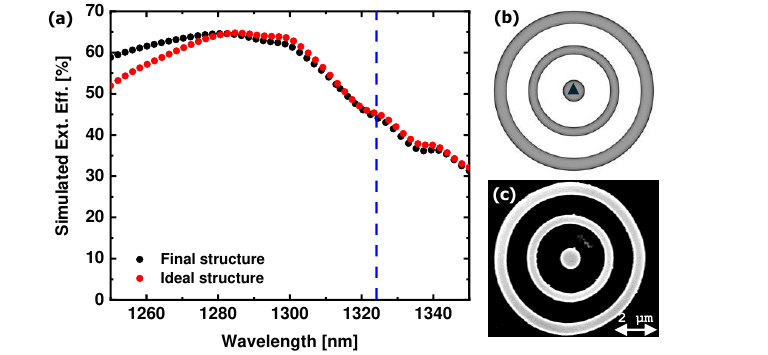}  
	\caption{(a) Simulated PEE as function of emission wavelength. Results for an idealized structure, depicted in panel (b), where the QD is positioned at the center of the mesa are plotted in red. For comparison, the black data points show simulation results taking the geometry of the actual fabricated structure, depicted shown in panel (c) into account, incorporating a QD with a lateral offset of 20 nm. Both simulations evaluate the extraction efficiency as a function of the emission wavelength of the integrated emitter. The blue dashed line at 1324~nm marks the X$^{\pm}$ emission at 4~K that was used to determine the experimental PEE.}
	\label{Fig:PEE_Sim}
\end{figure*}

Before the growth of the semiconductor heterostructures, both the vertical layer sequence and the circular Bragg grating (CBG) design were optimized to maximize the photon extraction efficiency (PEE) at a target wavelength range of 1280 to 1300~nm. This optimization was performed using the finite element method (FEM) solver \textsc{JCMsuite}, which enables rigorous numerical modeling of light emission in nanophotonic devices\cite{Rickert2019, jcmwave2025}. The simulations were carried out in a two-dimensional configuration by exploiting the symmetry of the CBG geometry. The refractive index dispersion and the position of the quantum dot (QD), represented as an electric dipole source, were also taken into account. The geometric configuration yielding the highest outcoupling efficiency into the collection optics with a numerical aperture (NA) of 0.8 was identified by varying design parameters such as the grating period, ring thickness, etch depth, and mesa diameter.

\begin{table*}
\centering
\begin{tabular}{|l|c|c|c|c|c|c|}
\hline
Design & Mesa Diameter (nm) & Ring 1 Width (nm) & Gap 1 Width (nm) & Ring 2 Width (nm) & Gap 2 Width (nm) & PEE$_\text{max}$ (\%) \\
\hline
Ideal structure & 891 & 354 & 1064 & 479 & 929 & 45  \\

Final structure & 885 & 390 & 1025 & 540 & 860 & $(43~\pm~1)$  \\
\hline
\end{tabular}
\caption{Simulated ideal geometrical parameters of the optimized CBG device for the 1280–1300~nm wavelength range, shown together with the realized structure.}
\label{tab:CBG_parameters}
\end{table*}

The results of the optimization procedure are summarized in Table~\ref{tab:CBG_parameters} and Fig.~\ref{Fig:PEE_Sim}. The table compares the target (ideal) CBG geometry used in the simulations with the final, fabricated structure, highlighting differences in mesa diameter, ring widths, and gap widths. Quantitatively, the mesa diameter decreases slightly from 891~nm to 885~nm (-0.7\%), Ring~1 width increases from 354~nm to 390~nm (+10\%), Gap~1 decreases from 1064~nm to 1025~nm (-3.7\%), Ring~2 width increases from 479~nm to 540~nm (+12.7\%), and Gap~2 decreases from 929~nm to 860~nm (-7.4\%). Despite these deviations, the maximum PEE only drops slightly from 45\% to $(43 \pm 1)\%$, indicating that the design is robust against the observed fabrication variations.

Figure~\ref{Fig:PEE_Sim}(a) shows the simulated wavelength-dependent PEE for the ideal structure (red data points) and the fabricated device (black data points). Finite-element method (FEM) simulations of the realized device, which incorporate the measured geometrical parameters of the fabricated structure as well as the spatial offset of the emitter within the central mesa, predict a PEE of $(43 \pm 1)\%$ at an emission wavelength of approximately 1324~nm. This demonstrates good agreement between theory and experiment while explicitly accounting for deviations from the target geometry introduced during fabrication.

\section{EPITAXIAL GROWTH}

\begin{figure*}[t]  
	\centering
	\includegraphics[width=0.85\textwidth]{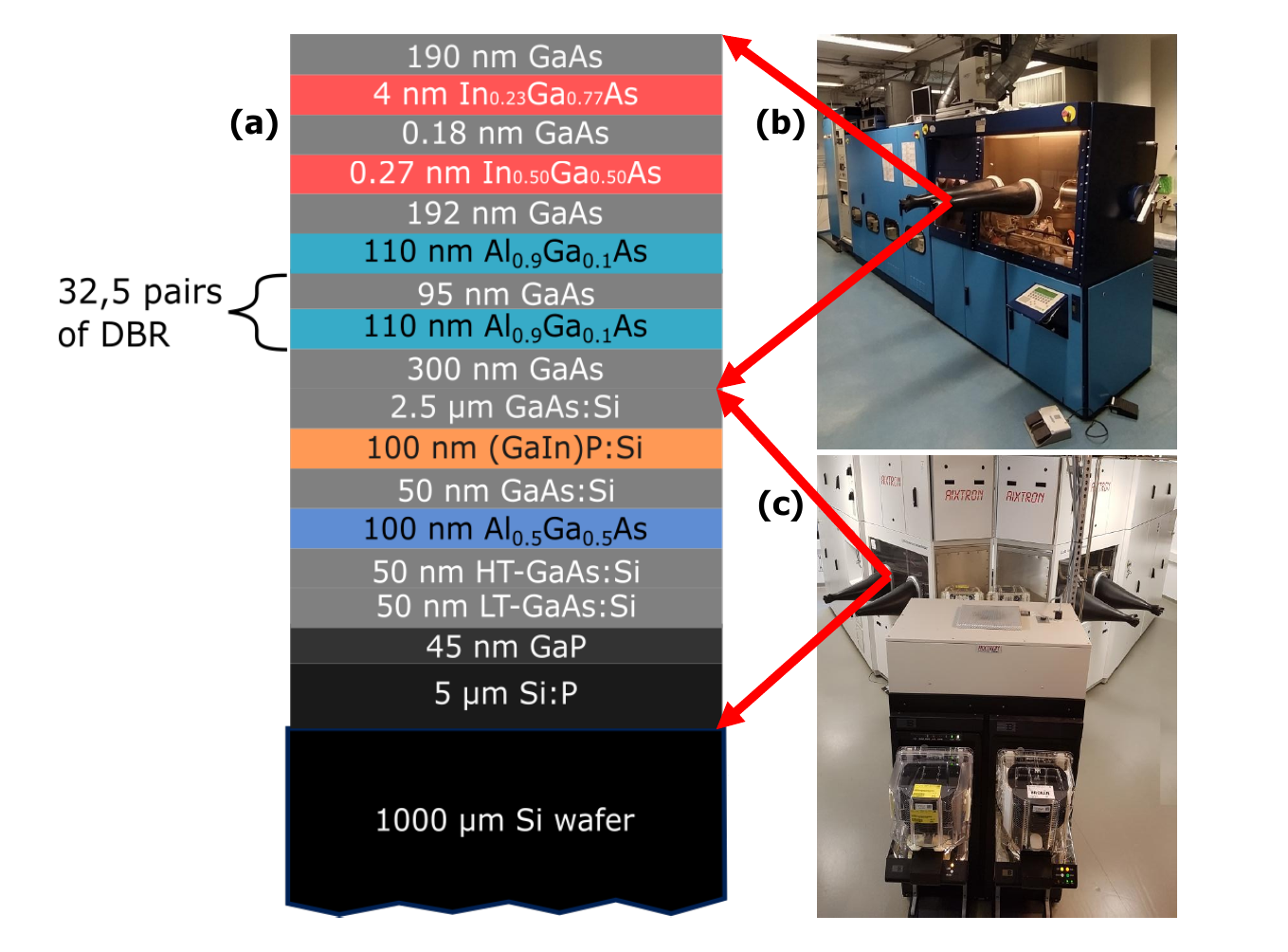}  
	\caption{(a) Schematic layer structure of the investigated sample, grown on a 1 mm-thick undoped silicon (Si) wafer. The lower part of the structure, including a 45 nm GaP buffer layer as well as Al$_{0.5}$Ga$_{0.5}$As and GaInP layers, was grown using a 300 mm \textit{Crius} cluster tool (Aixtron AG), as shown in (c), at the University of Marburg. The upper section of the structure was fabricated at the Technical University of Berlin using a III/V-MOVPE system (Aixtron AG, 200/4), see (b). This includes 32.5 mirror pairs of GaAs/Al$_{0.9}$Ga$_{0.1}$As distributed Bragg reflector (DBR), designed to enhance the PEE of the embedded QDs. The DBR layers were deposited at 700°C. A 0.27 nm-thin In$_{0.5}$Ga$_{0.5}$As wetting layer was subsequently grown at a lower temperature of 500°C, followed by a 4 nm In$_{0.5}$Ga$_{0.5}$As strain-reducing layer. The structure was capped with GaAs to complete a $\lambda$-cavity.}
	\label{Fig:Structure}
\end{figure*}

The investigated QD heterostructure was fabricated in a two-step epitaxial growth process, with the two steps performed in a separate metal--organic chemical vapor deposition (MOCVD) reactor. A schematic overview of the full layer structure, along with images of the respective growth systems---namely the 300~mm Crius R CCS Cluster (Aixtron AG) used for the initial Si/GaP template fabrication and the AIX 200/4 (Aixtron AG) used for QD growth, ---are presented in Fig.~\ref{Fig:Structure}(a), (b), and (c), respectively. The initial heteroepitaxial template was grown by NAsP III/V GmbH at the Philipps-University of Marburg using a twin-reactor 300 mm Crius R CCS Cluster MOCVD system. The substrate consisted of a 300~mm Si (001) wafer with exact orientation ($\pm 0.5^\circ$), on which a 5~\textmu m-thick Si:P buffer was first deposited as part of an established GaP-on-Si process.\cite{Volz2011} This was followed by the growth of a GaP nucleation layer and a relaxation buffer composed of AlGaAs/GaAs to manage lattice mismatch and minimize threading dislocations.\cite{Jung2017b} Following the initial growth step, the 300~mm wafer was cleaved into 4~cm~$\times$~4~cm dies and transferred to the AIX 200/4 reactor at the Technical University of Berlin for QD epitaxy. Prior to growth, the samples were thermally deoxidized at 735~$^\circ$C to remove any native oxide from the surface. A 300~nm undoped GaAs buffer layer was then deposited at high V/III ratio ($\approx 200$) to ensure high crystal quality. Subsequently, a DBR comprising 33.5 pairs of quarter-wavelength ($\lambda/4$) thick GaAs (95~nm) and Al$_{0.9}$Ga$_{0.1}$As (110~nm) layers was grown. This DBR serves as a backside mirror in conjunction with a later-fabricated CBG structure to enhance photon extraction from the embedded QDs. After the DBR, a 192~nm GaAs spacer layer was grown, forming half of the optical cavity. The substrate temperature was then reduced to 500~$^\circ$C for QD formation. A thin In$_{0.5}$Ga$_{0.5}$As wetting layer (0.27~nm) was deposited, followed by a growth interruption of approximately 60~seconds to allow QD nucleation. QDs with a density of $\sim (1-5 \times 10^9)$~cm$^{-2}$ were then capped with a 0.18~nm GaAs layer and a 4~nm strain reducing layer (SRL) of In$_{0.23}$Ga$_{0.77}$As. The SRL mitigates lattice strain imposed by the GaAs matrix, enabling the formation of larger QDs with redshifted emission in the telecom O-band.\cite{Srocka2020-ms} To remove residual indium from the surface, the structure was annealed at 615~$^\circ$C. Finally, a 190~nm GaAs layer was deposited to complete the $\lambda/4$ cavity.

\section{IN-SITU EBL INTEGRATION}

\begin{figure*}[t]  
	\centering
	\includegraphics[width=\textwidth]{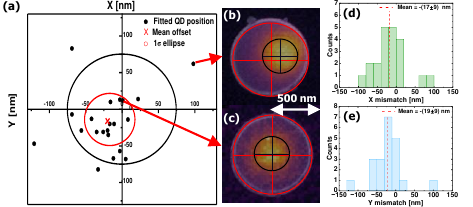}  
	\caption{(a) Spatial distribution of the extracted QD positions with respect to the central mesa's center of the 21 fabricated structures. The positions were extracted by CL mapping, and two representative examples are highlighted: one with a lateral [x,y] offset of [98, 62]~nm (b) and another with an offset of [$-3$, 13]~nm (c) relative to the mesa center. (d) Histogram of the X-coordinate mismatch, evaluated with an optimal bin width of $\approx 18$~nm according to the Freedman--Diaconis rule. The distribution is centered at an average offset of $-(17 \pm 9)$~nm with a standard deviation (spread) of $(46 \pm 10)$~nm, showing that the QDs exhibit a slight systematic shift to the negative X direction. (e) Histogram of the Y-coordinate mismatch, obtained with the same binning approach, yields an average offset of $-(19 \pm 9)$~nm and a spread of $(41 \pm 8)$~nm, confirming a comparable localization accuracy along the Y axis.}
	\label{Fig:Intgra}

\end{figure*}

For the in-situ integration of QDs into the CBG structures, the sample was first spin-coated with the E-beam resist AR-P~6200-13 (CSAR~62, Allresist GmbH) at 4000~rpm, resulting in a resist thickness of 400~nm . Cathodoluminescence (CL) mapping was then performed at 20~K to locate optically active QDs with an integration time of 80~ms per 500~nm pixel, corresponding to an electron-beam dose of 7.7~mC/cm$^2$. This dose was chosen as a compromise between achieving sufficient signal for emitter localization and preventing under- or over-exposure of the resist during the subsequent in-situ electron-beam lithography (EBL) step. After recording the luminescence map, the emitter coordinates were extracted and used to generate CBG mask layouts aligned to the selected QDs. EBL was performed at 20~K with proximity-effect–corrected patterns and high exposure doses in the negative-tone regime (0.2–40~mC/cm$^2$). After development, the patterned resist served as an etch mask for the final structure transfer via inductively coupled plasma reactive-ion etching (ICP-RIE). Further details of the deterministic nanophotonic device fabrication platform can be found in Ref.~\cite{Rodt2021}. A major challenge of in-situ QD integration, particularly for emitters in the telecom bands, is the long CL mapping time required due to the reduced signal-to-noise ratio (SNR) of InGaAs detectors compared to Si CCDs used for 780–900~nm QDs. To mitigate this and reduce mapping duration (and hence cryostat drift), machine-learning–based image processing may be employed \cite{Donges2022, Madigawa2024}. ML models trained for pattern recognition on low-SNR luminescence maps can enhance emitter localization by denoising and accurately fitting the emission spots.

After fabrication, the positions of QDs within the central mesa were quantitatively analyzed via post-integration CL mapping. The offsets of a total of 21 QD-CBGs are presented in Fig.~\ref{Fig:Intgra}. Along the $X$ axis, individual displacements range from approximately $-121$~nm to $+98$~nm, with a mean value of $\langle X \rangle = -17 \pm 9$~nm and a standard deviation of $\sigma_X = 42 \pm 7$~nm. Along the $Y$ axis, displacements range from $-82$~nm to $+83$~nm, with a mean of $\langle Y \rangle = -19 \pm 9$~nm and $\sigma_Y = 40 \pm 6$~nm. These results indicate a slight but systematic offset of the QD positions toward negative $X$ and $Y$, likely caused by thermal drift of the cryostat. The mean radial displacement from the mesa center is $\langle r \rangle = 53 \pm 7$~nm, with a corresponding radial standard deviation of $\sigma_r = 35 \pm 6$~nm. Histograms of the QD positions along $X$ and $Y$ (Fig.~\ref{Fig:Intgra}d, e) show approximately Gaussian distributions, slightly offset from the mesa center, reflecting the statistical spread of the emitter localization.

The statistical evaluation of QD positions highlights the overall accuracy and reproducibility of the in-situ EBL process employed for device fabrication. The observed offsets of approximately -20~nm in both the X and Y directions represent a systematic deviation that can be related to a thermal drift of the cryostat. Importantly, the spatial spreads of about 40~nm indicate that the alignment accuracy of the process remains well within the sub-100-nm range suitable for CBG integration.

\section{OPTICAL RESULTS}

\begin{figure*}[t]  
	\centering
	\includegraphics[width=0.8\textwidth]{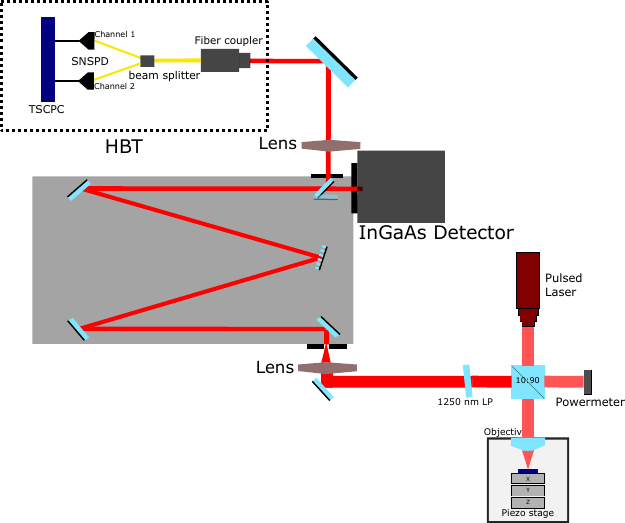}  
	\caption{A schematic of the \textmu PL setup used to characterize the optical and time-resolved PL properties of the investigated QD-CBG device on the GaP/Si substrate.}
	\label{Fig:Setup}
\end{figure*}

\begin{figure*}[t]  
	\centering
	\includegraphics[width=1\textwidth]{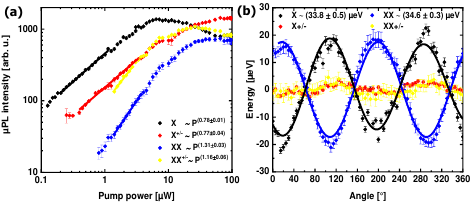}  
	\caption{(a) Double-logarithmic plot of the micro-\textmu PL intensity as a function of excitation power, demonstrating the power dependence of the excitonic transitions: neutral exciton (X, black data points), charged exciton (X$^{\pm}$, red), biexciton (XX, blue), and charged biexciton (XX$^{\pm}$, yellow). The solid lines represent fits to the data, with slopes corresponding to the respective emission processes. (b) Results of polarization-resolved \textmu PL measurements of the same emission lines reveal a fine structure splitting (FSS) of $\Delta_\text{FSS} = (34.2 \pm 0.4)\ \textmu \text{eV}$.}
	\label{Fig:PowPol}
\end{figure*}

\begin{figure*}[t]  
	\centering
	\includegraphics[width=\textwidth]{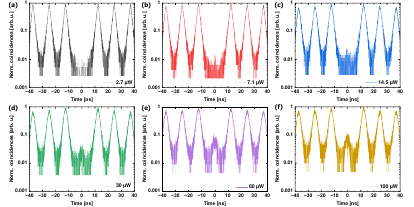}  
	\caption{Half-logarithmic presentation of the second-order photon autocorrelation function \( g^{(2)}(\tau) \) shown in linear scale in Fig. 3(b), measured for the charged-exciton emission line over a range of excitation powers: 2.7~µW, 7.1~µW, 14.5~µW, 30~µW, 60~µW, and 100~µW, shown in a-f, respectively. As the excitation power increases, a gradual rise in the multiphoton emission probability is observed, reflected in increasing \( g^{(2)}(0) \) values. Specifically, the measured \( g^{(2)}(0) \) are \( (0.007 \pm 0.001) \), \( (0.013 \pm 0.001) \), \( (0.027 \pm 0.001) \), \( (0.059 \pm 0.001) \), \( (0.134 \pm 0.003) \), and \( (0.172 \pm 0.004) \), respectively.}
	\label{Fig:g2s}
\end{figure*}

Optical characterization of the investigated device was performed using a micro-photoluminescence (\textmu PL) setup, schematically illustrated in Fig.~\ref{Fig:Setup}. The \textmu PL system consists of a closed-cycle cryostat equipped with a three-axis piezoelectric nanopositioning stage, enabling precise spatial alignment of the beam and structure. Excitation and collection were carried out through a high-numerical-aperture objective lens (NA = 0.82), ensuring efficient light coupling and high spatial resolution. The excitation source was an 80~MHz optical parametric oscillator (OPO) providing wavelength-tunable pulsed laser excitation, which was directed to the sample via a 10:90 (excitation:detection) beam splitter (BS). The PL signal was dispersed by a monochromator equipped with 150 and 900~lines/mm gratings, providing spectral resolutions of 150 and 20~\textmu eV, respectively, and detected using an InGaAs line detector. For second-order correlation measurements, the exit mirror of the monochromator was flipped to direct the emission into a Hanbury Brown and Twiss (HBT) interferometer. The HBT setup employed a single-mode fiber BS feeding two independent channels of superconducting nanowire single-photon detectors (SNSPDs) with a time resolution of 30 ps, while time-correlated single-photon counting (TCSPC) electronics were used to record the coincidence histogram and determine the second-order autocorrelation function, $g^{(2)}(\tau)$. Note that all values of $g^{(2)}(0)$ were obtained by dividing the area of the coincidence peak at $\tau = 0$ by the average area of the side peaks at $\tau \neq 0$.

To determine the origin of the observed emission lines, power-dependent and polarization-resolved \textmu PL measurements were performed (Figs. \ref{Fig:PowPol}(a) and (b)). Four transitions were identified: the neutral exciton (X), charged excitons ($X^{+/-}$), the neutral biexciton (XX), and charged biexcitons (XX$^{+/-}$). The X transition, centered at 0.9369~eV (1323.3 nm), exhibits a near-linear power dependence with an intensity scaling of $P^{(0.78 \pm 0.01)}$ (Fig. \ref {Fig:PowPol}(a)). The fine-structure splitting (FSS) of the X and XX states was extracted as (34.2~$\pm$0.4),\textmu eV from sinusoidal fits of the polarization-dependent peak energies (Fig. \ref{Fig:PowPol}(b), black and blue lines). The XX emission at 0.9346~eV (1326.5 nm) is characterized by orthogonal linear polarization relative to X and a super-linear power dependence of $P^{(1.31 \pm 0.03)}$, consistent with a biexciton–exciton cascade. Charged excitonic and biexcitonic states show no resolvable FSS and display linear or super-linear power dependencies, respectively, in agreement with their single- or double-charge character. These observations align with established optical signatures of excitonic complexes in InGaAs QDs.

The second-order photon autocorrelation function $g^{(2)}(\tau)$ of the charged-exciton (X$^{\pm}$) emission line measured for excitation powers of 2.7~$\mu$W, 7.1~$\mu$W, 14.5~$\mu$W, 30~$\mu$W, 60~$\mu$W, and 100~$\mu$W. In Fig.~\ref{Fig:g2s}(a--f), the data are presented in a half-logarithmic scale to better visualize the counts at $\tau = 0$.

\begin{figure*}[t]  
	\centering
	\includegraphics[width=\textwidth]{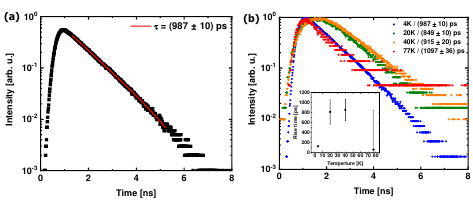}  
	\caption{(a) Time-resolved \textmu PL measurement of the X$^{+/-}$ transition (black data points) of the QD-CBG device under study. An exponential fit (red solid line) of the measured PL decay yields a lifetime of $(987 \pm 1)$~ps. (b) Temperature-dependent decay time of the excitonic (X) transition. An exponential fit is applied to extract the lifetime across the temperature series.}
	\label{Fig:TRPL}
\end{figure*}

Figure~\ref{Fig:TRPL}(a) presents the time-resolved \textmu 
PL measurement of the spectrally filtered exciton emission from the investigated QD under pulsed non-resonant excitation at 1240~nm and a temperature of 4~K. The decay dynamics were analyzed using a mono-exponential fit (red line) to the logarithmically plotted PL transient, yielding a radiative lifetime of $(987 \pm 10)$~ps. This lifetime is in excellent agreement with typical values reported for self-assembled InGaAs/GaAs QDs grown via the Stranski–Krastanov epitaxial mode on GaAs substrates.\cite{Groe2021, Maisch2024} The result confirms that the measured emission dynamics are predominantly governed by intrinsic excitonic recombination processes rather than by cavity-induced enhancement of the spontaneous emission rate.

Figure~\ref{Fig:TRPL}(b) shows the temperature-dependent time-resolved photoluminescence (TRPL) measurements recorded at 4, 20, 40, and 77~K. The exciton decay dynamics reveal that the radiative lifetime $t_1$ remains nearly constant at approximately 1~ns across the investigated temperature range, as shown in Fig.~\ref{Fig:TRPL}(b). In contrast, the extracted rise times $t_0$, shown in the inset of Fig.~\ref{Fig:TRPL}(b), which characterizes the carrier capture and thermalization processes from the GaAs barrier and InGaAs wetting layer into the QD ground state, exhibits a pronounced temperature dependence. At 4~K, a rise time of $t_0 = (126 \pm 7)$~ps is observed, indicative of a phonon bottleneck effect that limits carrier relaxation under non-resonant excitation conditions. As the temperature increases to 20~K and 40~K, $t_0$ increases markedly to approximately 800~ps, suggesting that enhanced carrier–phonon scattering and redistribution among higher-energy states slow down the relaxation into the QD ground state. Interestingly, at 77~K, the rise time decreases again to $t_0 = 61^{+78}_{-61}$ ps, pointing to a transition in the carrier relaxation mechanism. This non-monotonic trend likely arises from the activation of additional phonon-assisted scattering pathways at elevated temperatures, which facilitate faster thermalization and more efficient population of the QD states \cite{Heitz2001, Urayama2001}. Such behavior highlights the complex interplay between phonon-mediated relaxation and carrier recapture dynamics in self-assembled InGaAs QDs under varying thermal conditions.

\clearpage
\section*{REFERENCES}
	\bibliographystyle{unsrt}
	\bibliography{References}
\clearpage